\documentclass[a4paper,11pt]{article}
\usepackage{jheppub} 
\usepackage{lineno}

\usepackage{amsthm}
\usepackage{amssymb}
\usepackage{amsmath}
\usepackage{mathrsfs}
\usepackage{physics}
\usepackage{mathtools}
\usepackage{array}
\usepackage{booktabs}
\usepackage{enumitem}
\usepackage{MnSymbol}
\usepackage{tablefootnote}
\usepackage{setspace}
\usepackage{longtable}
\usepackage{needspace}
\usepackage{makeidx}

\theoremstyle{plain}

\theoremstyle{definition}

\theoremstyle{remark}

\allowdisplaybreaks

\title{\boldmath Coupled free fermion conformal field theories}

\author{Peter Bouwknegt}
\author{and Bolin Han}
\affiliation{Mathematical Sciences Institute, Australian National University,
Canberra, ACT 2601, Australia}

\emailAdd{peter.bouwknegt@anu.edu.au, bolin.han@anu.edu.au}

\abstract{
We study a specific class of CFTs that involve coupled free fermions, arising from parafermion 
CFTs and lattice constructions. We analyse their representation spaces and the underlying exclusion statistics of coupled free fermions using specific bases. 
In one particular case, we reveal an unexpected connection between the coset construction of parafermions,  the lattice construction, and and orbifold thereof. 
This connection is supported by proving a range of character identities within the context of coupled free fermions. 
Simultaneously, we obtain explicit expressions of certain string functions in terms of Dedekind eta functions.}

\begin{document}
\maketitle
\flushbottom

\section{Introduction}
\label{sec:intro}

The study of formal properties of CFTs was initiated in the late 1960s 
\cite{MACK1969174,polyakov1974nonhamiltonian,Ferrara:1971vh,Parisi:1972zm,Caianiello1978NonPerturbativeMI}. 
Early work was done in general dimension $d\ge 2$, where the Lie algebra of the group of conformal transformations is 
almost always finite-dimensional, except when $d=2$ where the group is infinite-dimensional. The infinite-dimensional 
symmetry group in 2D CFTs allows us to define the theories in a more abstract way via operator algebras and their representation theory.

With continuous developments since then, CFT has become an active area of research beyond its origins in 
statistical physics and attracted much attention for its intrinsic mathematical interest. 
The quest to classify and solve CFTs is a major goal of modern theoretical physics. 
A complete understanding of the space of CFTs is not yet available. It is only the rational CFT subclass that is reasonably well-understood.

Starting from the minimal models \cite{FQSminmodel,Belavin:1984vu},  various other constructions of rational CFTs, including the lattice 
construction \cite{DONG1993245}, WZW models \cite{KZWZW,Witten1984WZW,GepnerWitten}, GKO coset construction \cite{GKOcoset}, 
and the orbifold construction \cite{Dixon1985strorb,Dixon1988strorb,dixon1987orb,DGM1990orb,Dolan1996}, have been intensively investigated. 
Those theories that contain fields with spin-1 and spin-2 have drawn more attention from physicists, while the theories that contain fields of fractional spin, 
such as parafermionic CFTs, have not yet been fully explored.

Notable families of parafermion CFTs includes the FKF-type parafermions \cite{FRADKIN19801}, and the 
ZF-type parafermions, which relate to the WZW models as cosets in the form of $\widehat{\mathfrak{sl}(2)}_k/\widehat{\mathfrak{u}(1)}$ \cite{ZF1985}. 
A more general version of coset parafermion CFTs was introduced by Gepner in \cite{GEPNER87}, which can be thought of as cosets of WZW models in
the form of $\hat{\mathfrak{g}}_k/\widehat{\mathfrak{u}(1)}^n$, where $n$ is the rank of $\mathfrak{g}$, and $k$ to so-called level \cite{GEPNER89,Bratchikov_2000}.

We notice that when $\mathfrak{g}$ is simply-laced and $k=2$, Gepner's parafermions are spin-$\frac{1}{2}$ free fermions. 
Moreover, when $\mathfrak{g}\neq\widehat{\mathfrak{sl}(2)}$, these fermions are coupled to each other. 

In Section 2 we study the coupled free fermions arising from the coset CFT $\hat{\mathfrak{g}}_k/\widehat{\mathfrak{u}(1)}^n$, in particular for $\mathfrak g = \mathfrak{sl}(3)$ and 
$\mathfrak g = \mathfrak{sl}(4)$.  We construct a basis for the modules of this CFT in terms of the coupled free fermions so that their characters can be computed in terms 
of so-called universal chiral partition functions (`fermionic partition functions'), mathematically giving alternate explicit expressions for the string functions of 
affine Lie algebra modules.
In Section 3 we give another construction of coupled free fermions, associated to (rescaled) lattice CFTs, and observe an intriguing relation between coset construction for 
$\mathfrak{sl}(4)$ and the lattice CFT for $\mathfrak{sl}(3)$

Some details regarding the proof for the bases are given in Appendix A, while the proofs of the character identities can be found in a companion paper \cite{BCH}.



\section{The coset construction}

Given an affine Lie algebra $\hat{\mathfrak{g}}$ at level $k$, one can construct a parafermionic conformal field theory 
containing fields which are non-local,  of fractional spin, and chiral, as described in \cite{GEPNER87}. 
For simplicity, we only focus on the holomorphic part and assume that $\mathfrak{g}$ is simply-laced.

We associate a parafermionic field $\psi^{(\alpha)}$ to each $\alpha\in Q$, where $Q$ is the root lattice of $\mathfrak{g}$, 
and identify $\psi^{(\alpha)}$ and $\psi^{(\beta)}$ if $\alpha\equiv\beta\mod kQ$. 
Among these parafermionic fields, we are particularly interested in the so-called generating 
parafermions $\{\psi^{(\alpha)}\mid \alpha\in\Delta\}$, where $\Delta$ is the set of roots of $\mathfrak{g}$. 
The OPEs of generating parafermions are determined so that the following fields
\begin{align*}
    &J^{\alpha}(z)=\sqrt{\dfrac{2k}{\alpha^2}}c_\alpha\psi_{\alpha}e^{i\alpha\cdot\phi(z)/\sqrt{k}},\;\;\alpha\in \Delta,
    &J^{i}(z)=\dfrac{2i\sqrt{k}}{\alpha_i^2}\alpha_i\cdot\partial_z\phi(z)\;,\;\;\alpha_i\in\Delta_0,
\end{align*}
where $c_\alpha$ are cocycles, $\phi(z)$ is the vector of $n$ free bosons and $\Delta_0$ is the set of simple roots of $\mathfrak{g}$, 
obey the current algebra commutation relations 
\[J^{a}(z)J^b(w)\sim\dfrac{k\kappa^{ab}}{(z-w)^2}+\dfrac{f^{ab}{}_c\, J^c(w)}{z-w},\]
where $J^a(z)$ stands for either $J^{\alpha}(z)$ or $J^{i}(z)$ defined above, $\kappa^{ab}$ is the Killing form and $f^{ab}{}_c$ are structure constants of $\mathfrak{g}$ 
in the Chevalley basis.

It turns out that the cocycles and OPEs are given by
\begin{align}
    c_{\alpha}c_{-\alpha}=1,\;\;\;\;c_{\alpha}c_{\beta}=S_{\alpha,\beta}c_{\alpha+\beta},\notag\\
    \psi^{(\alpha)}(z)\psi^{(-\alpha)}(w)\sim\dfrac{1}{(z-w)^{2-|\alpha|^2/k}},\label{eq:paraOPE-1}\\
     \psi^{(\alpha)}(z)\psi^{(\beta)}(w)\sim\dfrac{c_{\alpha,\beta}\psi^{(\alpha+\beta)}(w)}{(z-w)^{1+(\alpha,\beta)/k}},\label{eq:paraOPE-2}
\end{align}
where $S_{\alpha,\beta}$ are some roots of unity and $c_{\alpha,\beta}$ are some constants to be fixed by the Borcherds identity \eqref{eq:paraBorId}.

Denote the energy-momentum tensor of the current algebra in the Sugawara form by $L^{(c)}(z)$, 
and that of the (uncoupled free) bosonic systems by $L^{(b)}(z)$. The energy-momentum tensor $L(z)$ of the parafermionic system is then $L(z)=L^{(c)}(z)-L^{(b)}(z)$
and the central charge of the parafermionic system is
\begin{equation}\label{eq:PFcc}
    c\left(\dfrac{\hat{\mathfrak{g}}_k}{\widehat{\mathfrak{u}(1)}^n}\right)= c\left(\hat{\mathfrak{g}}_k\right)-c\left(\widehat{\mathfrak{u}(1)}^n\right)=\dfrac{k\dim\mathfrak{g}}{k+h^{\vee}}-n.
\end{equation}
where $h^{\vee}$ is the dual Coxeter number of $\mathfrak g$.

The conformal dimension of a parafermionic field $\psi^{(\alpha)}$ with respect to $L(z)$ is given by
\begin{equation}\label{eq:genPFcfwt}
    h(\alpha)=-\dfrac{|\alpha|^2}{2k}+n(\alpha),
\end{equation}
where $n(\alpha)$ is the minimal number of roots of $\mathfrak{g}$ from which $\alpha$ is composed. The mode expansions of parafermionic fields $\psi^{(\alpha)}$ are of the form 
\begin{equation}\label{eq:paramode}
    \psi^{(\alpha)}(z)=\sum_{n\in \mathbb{Z}+\epsilon(\alpha)} \psi^{(\alpha)}_n z^{-n-h(\alpha)},
\end{equation}
where the twisting $\epsilon\in\mathbb{R}/\mathbb{Z}$ depends on relevant OPEs. Unitarity is defined by the conjugation relation
\[\left(\psi^{(\alpha)}_n\right)^{\dag}=\psi^{(\alpha)}_{-n},\]
which is consistent with the generalised commutation relations \eqref{eq:gcommrel}. 

One can derive generalised commutation relations between the modes of parafermionic fields from the Borcherds identity for 
parafermionic fields
\begin{align}\label{eq:paraBorId}
  &\sum_{j\ge0}\binom{m}{j}[[AB]_{n+1+j}C]_{m+k+1-j} \notag\\
   =&(-1)^j\binom{n}{j}\sum_{j\ge0}[A[BC]_{k+1+j}]_{m+n+1-j}-\mu_{AB}(-1)^{\alpha_{AB}-n}[B[AC]_{m+1+j}]_{n+k+1-j},
\end{align}
where $n\equiv\alpha_{AB}\mod\mathbb{Z}$, $ m\equiv\alpha_{AC}\mod\mathbb{Z}$, $k\equiv\alpha_{BC}\mod\mathbb{Z}$ 
and the commutation factors $\mu_{AB}$ and singularity $\alpha_{AB}$ are given by the axiom
\begin{equation}\label{eq:OPEcomm}
A(z)B(w)(z-w)^{\alpha_{AB}}=\mu_{AB}B(w)A(z)(w-z)^{\alpha_{AB}}.
\end{equation}

The parafermionic primary fields of $\hat{\mathfrak{g}}_k/\widehat{\mathfrak{u}(1)}^n$, denoted by $\Phi^{\Lambda}_{\lambda}$, 
are labelled by $\Lambda\in \hat{P}^{(k)}_+$, a dominant integral weight at level $k$, and an integral weight 
$\lambda\in \hat{P}$ such that $\Lambda-\lambda\in\hat{Q}$, where $\hat{Q}$ is the root lattice of $\hat{\mathfrak{g}}$. 
We notice that $\Phi^{\Lambda}_{\lambda}$ and $\Phi^{\Lambda'}_{\lambda'}$ are 
identified if $$\Lambda'=\Lambda\;\;\text{and}\;\;\lambda'\equiv \lambda\mod k\hat{Q}$$ or $$\Lambda'=\sigma\Lambda\;\;\text{and}\;\;\lambda'=\sigma\lambda,$$
where $\sigma\in\text{Aut}(\text{Dyn}(\mathfrak{g}))$. The modules of $\hat{\mathfrak{g}}_k/\widehat{\mathfrak{u}(1)}^n$, denoted by $\mathcal{L}^{\Lambda}_{\lambda}$, 
are therefore labelled and identified in the same manner.

The generating parafermions $\psi^{(\alpha)}$ act on modules $\mathcal{L}^{\Lambda}_{\lambda}$ as chiral vertex operators (CVOs). 
Consider a generic CVO $$\phi^{\Lambda^{(i)}}_{\lambda^{(i)}}\binom{i}{k\;j}(z):\mathcal{L}^{\Lambda^{(j)}}_{\lambda^{(j)}}\to \mathcal{L}^{\Lambda^{(k)}}_{\lambda^{(k)}},$$
whose mode expansion is 
\begin{equation}\label{eq:CVOexp}
    \phi^{\Lambda^{(i)}}_{\lambda^{(i)}}\binom{i}{k\;j}(z)=\sum_{n\in\mathbb{Z}}\phi^{\Lambda^{(i)}}_{\lambda^{(i)}}
    \binom{i}{k\;j}_{n-(h^{\Lambda^{(k)}}_{\lambda^{(k)}}-h^{\Lambda^{(j)}}_{\lambda^{(j)}})}z^{-n+(h^{\Lambda^{(k)}}_{\lambda^{(k)}}
    -h^{\Lambda^{(j)}}_{\lambda^{(j)}}-h^{\Lambda^{(i)}}_{\lambda^{(i)}})},
\end{equation}
where $h^{\Lambda}_{\lambda}$ is the conformal dimension of the field $\Phi^{\Lambda}_{\lambda}$, given by
\begin{equation}\label{eq:PFcfwt}
    h^{\Lambda}_{\lambda}=\dfrac{(\Lambda,\Lambda+2\rho)}{2(k+h^\vee)}-\dfrac{|\lambda|^2}{2k}+n^{\Lambda}_{\lambda},
\end{equation}
where $n^{\Lambda}_{\lambda}$ is some integer.
(If $\Lambda=k\Lambda_0$, then $n^{\Lambda}_{\lambda}$ may be described as the minimal number of finite roots from which the 
finite part of $\lambda$ is composed. If $\lambda$ is a weight in the representation $\Lambda$, then $n^{\Lambda}_{\lambda}=0$.)

The number of such CVOs are determined by the fusion rules of $\hat{\mathfrak{g}}_k/\widehat{\mathfrak{u}(1)}^n$, which are given  by 
\begin{equation}\label{eq:parafus}
    \Phi^{\Lambda^{(i)}}_{\lambda^{(i)}}\times \Phi^{\Lambda^{(j)}}_{\lambda^{(j)}}=\sum_k \mathcal{N}_{ij}^k\, \Phi^{\Lambda^{(k)}}_{\lambda^{(i)}+\lambda^{(j)}\mod k\hat{Q}}
\end{equation}
where $\mathcal{N}_{ij}^k$ are fusion rules of $\hat{\mathfrak{g}}_k$. 


The characters of $\mathcal{L}^{\Lambda}_{\lambda}$, denoted by $b^{\Lambda}_{\lambda}$, are defined in the standard way, that is,
\[b^{\Lambda}_{\lambda}(\tau):=\text{Tr}_{\big| {\mathcal{L}^{\Lambda}_{\lambda}}}  \ q^{\left(L_0-c/24\right)}\,, \qquad q=e^{2\pi i\tau} \]
where $L_0$ is the zeroth mode of the energy-momentum tensor $L(z)$ of the parafermionic system. Recalling that $L(z)=L^{(c)}(z)-L^{(b)}(z)$, 
we have, as shown in \cite{GEPNER87},
\begin{equation}\label{eq:cosetstrfun}
    b^{\Lambda}_{\lambda}(\tau)=\dfrac{\text{Tr}_{\big|\mathcal{L}^{\Lambda}_{\lambda}}\;  q^{\left(L_0^{(c)}-c(\hat{\mathfrak{g}}_k)/24\right)}}
    {\text{Tr}_{\big| \mathcal{L}^{\Lambda}_{\lambda}}\; q^{\left(L_0^{(b)}-n/24\right)}}=\eta(\tau)^{n} c^{\Lambda}_{\lambda}(\tau),
\end{equation}
where $\eta(\tau)$ is the Dedekind eta function and $c^{\Lambda}_{\lambda}(\tau)$ are the string functions of $\hat{\mathfrak{g}}_k$ \cite{KACPETERSON}.

We notice that when $\mathfrak{g}$ is simply-laced and $k=2$, this construction is of particular interest because then the OPEs \eqref{eq:paraOPE-1} 
and \eqref{eq:paraOPE-2} for generating parafermions reduce to
\begin{align}
    &\psi^{(\alpha)}(z)\psi^{(-\alpha)}(w)\sim\dfrac{1}{(z-w)},\label{eq:freefer}\\
     &\psi^{(\alpha)}(z)\psi^{(\beta)}(w)\sim\dfrac{c_{\alpha,\beta}\psi^{(\alpha+\beta)}(w)}{(z-w)^{1/2}},\;\;\text{if}\;\alpha+\beta\in\Delta.\label{eq:couplefer}
\end{align}
Keeping in mind that $\alpha\equiv -\alpha\mod 2Q$ and the conformal dimension of any generating parafermion $\psi^{(\alpha)}$ is $\frac{1}{2}$ 
according to \eqref{eq:genPFcfwt}, we see that \eqref{eq:freefer} tells us that we have a set of real free fermions $\{\psi^{(\alpha)}\mid \alpha\in \Delta_+\}$ 
and \eqref{eq:couplefer} tells us that these fermions are coupled to each other according to the root structure of $\mathfrak{g}$. 
Therefore, we will call such a system a ``coupled free fermion CFT" hereafter.

It is not hard to see that when $\mathfrak{g}=\mathfrak{sl}_2$, the coupled free fermion CFT simply recovers the free fermion CFT.
Thus, coupled free fermion CFTs are natural generalisations of the free fermion CFT. 

For a coupled free fermion CFT $\widehat{\mathfrak{sl}(n+1)}_2/\widehat{\mathfrak{u}(1)}^{n}$, the twisting $\epsilon$ in the 
mode expansion \eqref{eq:paramode} can only be $0$ or $\frac{1}{2}$, corresponding to NS- or R-sector respectively. 
The generalised commutation relations\footnote{Also known as $Z$-algebra relations \cite{Lepowsky1984}.}  
between the modes of coupled free fermions derived from \eqref{eq:paraBorId} are given in \cite{DING1994,BorisPF} as follows:
\begin{equation}\label{eq:gcommrel}
    \begin{aligned}
    \psi^{(\alpha)}_n\psi^{(\alpha)}_m+\psi^{(\alpha)}_m\psi^{(\alpha)}_n&=\delta_{m+n,0},&\\[0.7ex]
    \psi^{(\alpha)}_n\psi^{(\beta)}_m+\mu_{\alpha,\beta}\psi^{(\beta)}_m\psi^{(\alpha)}_n&=0,&\text{if}\;\alpha+\beta\not\in\Delta,\\
    \sum_{l\ge0}\binom{l-\frac{1}{2}}{l}\left(\psi^{(\alpha)}_{m-\frac{1}{2}-l}\psi^{(\beta)}_{n+\frac{1}{2}+l}+
    \mu_{\alpha,\beta}\psi^{(\beta)}_{n-l}\psi^{(\alpha)}_{m+l}\right)&=c_{\alpha,\beta}\psi^{(\alpha+\beta)}_{m+n},&\text{if}\;\alpha+\beta\in\Delta,
    \end{aligned}
\end{equation}
where the commutator factors $\mu_{\alpha,\beta}$ are some roots of unity that satisfy 
\begin{align*}
    &\mu_{\alpha,\beta}\mu_{\beta,\alpha}=1, &\mu_{\alpha,\beta}c_{\alpha,\beta}=c_{\beta,\alpha}.
\end{align*}
It is also calculated in \cite{DING1994,BorisPF} using \eqref{eq:paraBorId} that
\begin{equation}\label{eq:CFFEM}
    L(z)=\frac{4}{n+3}\sum_{\alpha\in \Delta_+}L^{(\alpha)}(z),
\end{equation}
where $L^{(\alpha)}(z)$ is the energy-momentum tensor of a free fermion subalgebra:
$$L^{(\alpha)}(z):=-\frac{1}{2}:\psi^{(\alpha)}(z)\partial\psi^{(\alpha)}(z):.$$
The mode expansion of $L(z)$ is then given by
\[L(z)=\sum_{n}L_nz^{-n-2},\]
with
\[L_n=\frac{4}{n+3}\sum_{\alpha\in \Delta_+}L_n^{(\alpha)},\]
where
\[L_n^{(\alpha)}=	\left\{
	\begin{aligned}
	&\frac{1}{2}\sum_{r\in\mathbb{Z}+\frac{1}{2}}\left(r+\frac{1}{2}n\right):\psi^{(\alpha)}_{-r}\psi^{(\alpha)}_{n+r}:&(\text{NS}),\\
&\frac{1}{2}\sum_{r\in\mathbb{Z}}\left(r+\frac{1}{2}n\right):\psi^{(\alpha)}_{-r}\psi^{(\alpha)}_{n+r}:+\frac{1}{16}\delta_{n,0}&(\text{R}).
	\end{aligned}
	\right.\]
Here, whether the mode numbers are integers or half-integers is determined by \eqref{eq:CVOexp} depending on which particular intertwiner
we are considering.

It is easily checked that the modes of $L(z)$ form the Virasoro algebra with central charge
\begin{equation}\label{eq:CFFcc}
    c\left(\dfrac{\widehat{\mathfrak{sl}(n+1)}_2}{\widehat{\mathfrak{u}(1)}^{n}}\right)=\dfrac{n(n+1)}{n+3},
\end{equation}
which agrees with \eqref{eq:PFcc}. We notice that this number is the same as the sum of the first $n$ central charges of the discrete series of 
unitary minimal models, that is,
\begin{equation}\label{eq:CFFMINcc}
    c\left(\dfrac{\widehat{\mathfrak{sl}(n+1)}_2}{\widehat{\mathfrak{u}(1)}^{n}}\right)=\sum_{m=1}^{n} \left(1-\dfrac{6}{(m+2)(m+3)}\right).
\end{equation}
Indeed, it has been verified in \cite{BelGep} that the coset parafermionic CFT $\widehat{\mathfrak{sl}(n+1)}_2/\widehat{\mathfrak{u}(1)}^{n}$ 
can be decomposed into minimal models, taking these cosets as special cases of the AGT correspondence. 
Detailed analysis and examples of such decompositions can be found in \cite{BHphDthesis}.

In the case of $\widehat{\mathfrak{sl}(3)}_2/\widehat{\mathfrak{u}(1)}^2$, let $\alpha_1,\alpha_2$ denote the simple roots of $\mathfrak{sl}(3)$. 
Then the positive roots of $\mathfrak{sl}(3)$ are $\{\alpha_1,\alpha_2,\alpha_3:=\alpha_1+\alpha_2\}$ and we 
have 3 coupled free fermions $\psi^{(1)}:=\psi^{(\alpha_1)}, \psi^{(2)}:=\psi^{(\alpha_2)}$ and $\psi^{(3)}:=\psi^{(\alpha_3)}$. 
There are 8 inequivalent primary fields in this model. 

The algebraic structure of this model was studied in \cite{BorisPF}, where the author noted that a kind of
Poincar\'e-Birkhoff-Witt (PBW) theorem should hold for the model, but they failed to determine one. 
Despite the fact that, in \cite{Ard2002}, the author had found a basis involving (positive and negative) modes of 
parafermions associated with simple roots, with our understanding of coupled 
free fermions in terms of chiral vertex operators together with the explicit generalised commutation relations, 
we have found that it is straightforward to state a basis of PBW type in terms of non-positive modes of all coupled free fermions.

Let $\mathcal{B}_1$ be set of states in the form of
\begin{equation}\label{eq:sl3basis-123}
    \begin{aligned}
      \psi^{(3)}_{-N_3-\frac{N_1+N_2}{2}+\frac{1}{2}-s^{(3)}_{N_3}}\dots\psi^{(3)}_{-\frac{N_1+N_2}{2}-
      \frac{1}{2}-s^{(3)}_{1}}&\psi^{(2)}_{-N_2-\frac{N_1}{2}+\frac{1}{2}-s^{(2)}_{N_2}}\dots\\
    &\dots\psi^{(2)}_{-\frac{N_1}{2}-\frac{1}{2}-s^{(2)}_{1}}\psi^{(1)}_{-N_1+\frac{1}{2}-s^{(1)}_{N_1}}\dots\psi^{(1)}_{-\frac{1}{2}-s^{(1)}_{1}}\ket{0}
    \end{aligned}
\end{equation}
with $N_1,N_2,N_3\ge0$ and $s^{(i)}_N\ge\dots s^{(i)}_2\ge s^{(i)}_1\ge0$ for $i=1,2,3$.

Let $\mathcal{B}_2$ be the set of states in the form of 
\begin{equation}\label{eq:sl3basis-twi}
    \begin{aligned}
      \psi^{(3)}_{-N_3-\frac{N_1+N_2}{2}+1-s^{(3)}_{N_3}}\dots\psi^{(3)}_{-\frac{N_1+N_2}{2}-s^{(3)}_{1}}&\psi^{(2)}_{-N_2-\frac{N_1}{2}+1-s^{(2)}_{N_2}}\dots\\
    &\dots\psi^{(2)}_{-\frac{N_1}{2}-s^{(2)}_{1}}\psi^{(1)}_{-N_1+\frac{1}{2}-s^{(1)}_{N_1}}\dots\psi^{(1)}_{-\frac{1}{2}-s^{(1)}_{1}}\ket{1'}
    \end{aligned}
\end{equation}
with $N_1,N_2,N_3\ge0$ and $s^{(i)}_N\ge\dots s^{(i)}_2\ge s^{(i)}_1\ge0$ for $i=1,2,3$. 

We claim that $\mathcal{B}_1$ and $\mathcal{B}_2$ are the bases of the untwisted sector and twisted sector 
of $\widehat{\mathfrak{sl}(3)}_2/\widehat{\mathfrak{u}(1)}^2$ respectively. 
One can inductively prove both the independence and the spanning property of our 
proposed basis with extensive manipulations of generalised commutation relations \eqref{eq:gcommrel}. 
Instead, in this paper we prove the spanning property by an inductive proof (see Appendix \ref{app:sl3}) and
conclude the independence from the equality of dimensions.  The latter can be proven by a combinatorial argument as given in \cite{Ard2002} or 
by number theory techniques as given in \cite{BCH} using 
Universal Chiral Partition Functions (UCPFs) \cite{KKMM1992,KKMM1993,berkovich1999universal,BouwknegtUCPF}.

In the case of $\widehat{\mathfrak{sl}(4)}_2/\widehat{\mathfrak{u}(1)}^3$, let $\alpha_1,\alpha_2,\alpha_3$ denote 
the simple roots of $\mathfrak{sl}(4)$ such that $(\alpha_1,\alpha_3)=0$. Then the positive roots are 
$\Delta_+=\{\alpha_1,\alpha_2,\alpha_3,\alpha_4:=\alpha_1+\alpha_2,\alpha_5:=\alpha_2+\alpha_3,\alpha_6:=\alpha_1+\alpha_2+\alpha_3\}$ and therefore we have 6 coupled free fermions $\{\psi^{(i)}:=\psi^{(\alpha_i)}\mid \alpha_i\in\Delta_+\}$. There are 24 inequivalent primary fields in this model. 

The algebraic structure of this model is also briefly mentioned in \cite{BorisPF}.  Although the author noted that the commutation factors between
the orthogonal fermions are equal
$\pm 1$, they did not specify any constraint to determine the sign. We have found that the constraints for $\widehat{\mathfrak{sl}(4)}_2/\widehat{\mathfrak{u}(1)}^3$ are 
\begin{equation}\label{eq:sl4para-0}
    \mu_{54}=\dfrac{c_{14}c_{21}}{\mu_{26}c_{61}c_{15}}=\dfrac{c_{12}c_{64}}{c_{16}c_{25}},
\end{equation}
where $\mu_{ij}:=\mu_{\alpha_i,\alpha_j}$ and $c_{ij}:=c_{\alpha_i,\alpha_j}$. Explicitly, a compatible choice of parameters could be
\begin{equation}\label{eq:sl4para}
\begin{aligned}
    &c_{ij}=\mu_{ij}c_{ji}, \; \mu_{ij}\mu_{ji}=1,\\
    &\mu_{13}=\mu_{26}=1,\; \mu_{54}=-1, \\
    \mu_{12}=\mu_{24}=\mu_{41}=\mu_{15}=&\mu_{56}=\mu_{61}=\mu_{23}=\mu_{35}=\mu_{52}=\mu_{43}=\mu_{64}=\mu_{36}=x^2,\\
    c_{12}=c_{24}=c_{41}=c_{15}=&c_{56}=c_{61}=c_{23}=c_{35}=c_{52}=c_{43}=c_{64}=c_{36}=\dfrac{x}{\sqrt{2}}.
\end{aligned}
\end{equation}
where $x$ is again an 8th root of unity which is chosen to be $e^{-\frac{i\pi }{4}}$.

Let $\mathcal{B}_3$ be the set of states in the form of
\begin{equation}\label{eq:sl4basis}
    \begin{aligned}
      &\psi^{(6)}_{-N_6-\frac{N_1+N_3+N_4+N_5}{2}+\frac{1}{2}-s^{(6)}_{N_6}}\dots\psi^{(6)}_{-\frac{N_1+N_3+N_4+N_5}{2}
      -\frac{1}{2}-s^{(6)}_{1}}\psi^{(5)}_{-N_5-N_4-\frac{N_1+N_2+N_3}{2}+\frac{1}{2}-s^{(5)}_{N_5}}\dots\\&\dots\psi^{(5)}_{-N_4
      -\frac{N_1+N_2+N_3}{2}-\frac{1}{2}-s^{(5)}_{1}}\psi^{(4)}_{-N_4-\frac{N_1+N_2+N_3}{2}+\frac{1}{2}-s^{(4)}_{N_4}}
      \dots\psi^{(4)}_{-\frac{N_1+N_2+N_3}{2}-\frac{1}{2}-s^{(4)}_{1}}\psi^{(3)}_{-N_3-\frac{N_2}{2}+\frac{1}{2}-s^{(3)}_{N_3}}
      \dots\\&\dots\psi^{(3)}_{-\frac{N_2}{2}-\frac{1}{2}-s^{(3)}_{1}}\psi^{(2)}_{-N_2-\frac{N_1}{2}+\frac{1}{2}-s^{(2)}_{N_2}}
      \dots\psi^{(2)}_{-\frac{N_1}{2}-\frac{1}{2}-s^{(2)}_{1}}\psi^{(1)}_{-N_1+\frac{1}{2}-s^{(1)}_{N_1}}\dots\psi^{(1)}_{-\frac{1}{2}-s^{(1)}_{1}}\ket{0}
    \end{aligned}
\end{equation}
with $N_1,N_2,\dots,N_6\ge0$ and $s^{(i)}_N\ge\dots s^{(i)}_2\ge s^{(i)}_1\ge0$ for $i=1,2,\dots,6$.

We claim that $\mathcal{B}_3$ is a basis of PBW-type of the untwisted sector of $\widehat{\mathfrak{sl}(4)}_2/\widehat{\mathfrak{u}(1)}^3$. 
Analogous results for twisted sectors are included in \cite{BHphDthesis}. 
In order to prove these results, we need explicit expressions of the coset characters of $\widehat{\mathfrak{sl}(4)}_2/\widehat{\mathfrak{u}(1)}^3$ 
or, in other words, the string functions of $\widehat{\mathfrak{sl}(4)}_2$, according to \eqref{eq:cosetstrfun}. The expressions are as follows:
\begin{equation}\label{eq:sl4strfun}
    \arraycolsep=1.4pt\def\arraystretch{2.2}
    \begin{array}{lp{1.6cm}l}
         \multicolumn{3}{l}{b^{2000}_{2000}(\tau)+b^{2000}_{0020}(\tau)=\dfrac{\eta(4\tau)^4\eta(6\tau)^8}{\eta(\tau)\eta(2\tau)^4\eta(3\tau)^3
         \eta(12\tau)^4} +\dfrac{\eta(2\tau)^8\eta(3\tau)\eta(12\tau)^4}{\eta(\tau)^5\eta(4\tau)^4\eta(6\tau)^4},}\\
        b^{2000}_{2000}(\tau)-b^{2000}_{0020}(\tau) = \dfrac{\eta(\tau)^2}{\eta(2\tau)^2}, & & b^{2000}_{0101}(\tau) = \dfrac{\eta(2\tau)^2
        \eta(6\tau)^2}{\eta(\tau)^3\eta(3\tau)},\\
        b^{0101}_{2000}(\tau) = b^{0101}_{0002}(\tau) = \dfrac{3\eta(6\tau)^3}{\eta(\tau)^2\eta(2\tau)}, & & b^{0101}_{0101}(\tau) = 
        \dfrac{\eta(2\tau)^3\eta(3\tau)^2}{\eta(\tau)^4\eta(6\tau)},\\
        b^{1100}_{1100}(\tau)=\dfrac{\eta(4\tau)^5}{\eta(\tau)^3\eta(8\tau)^2}, & & b^{1100}_{0011}(\tau)=\dfrac{2\eta(2\tau)^2\eta(8\tau)^2}{\eta(\tau)^3\eta(4\tau)},
    \end{array}
\end{equation}
where affine weights are given in terms of Dynkin labels. 
The complete proofs of these expressions and the basis are given in \cite{BCH,BHphDthesis}.

\section{The lattice construction and its connection to the coset construction}

Motivated by recent investigations of the scaled lattice CFTs $V_{\sqrt{2}A_n}$ \cite{Bae2021,LAM2004614,KITAZUME2000893}, where $A_n$ denotes the root lattice
of $\mathfrak{sl}(n+1)$, we construct another family of coupled free fermions based on the lattice 
$\frac{1}{\sqrt{2}}A_n$.  In this CFT, each root of $A_n$ 
contributes a field of conformal dimension $\frac{1}{2}$, i.e.\ a fermion. For each $\alpha\in A_n$, define
\[\psi^{\alpha}(z)=:e^{i\frac{\alpha}{\sqrt{2}}\cdot\Phi(z)}:\]
and then the non-trivial OPEs between the generating fields $\psi^{\alpha}$ with $\alpha\in\Delta$ read as follows:
    \begin{align}
       \psi^{\alpha}(z)\overline{\psi^{\alpha}}(w)&\sim \dfrac{1}{z-w},\label{eq:latfreefer}\\
       \psi^{\alpha}(z)\psi^{\beta}(w)&\sim \dfrac{c_{\alpha,\beta}\psi^{\alpha+\beta}(w)}{(z-w)^{1/2}},\;\text{if}\;\;\alpha+\beta\in\Delta,\label{eq:latcpfer}
    \end{align}
where we used the fact that $\psi^{-\alpha}(z)=\overline{\psi^{\alpha}}(z)$. Now, \eqref{eq:latfreefer} 
suggests that each $\psi^{\alpha}$ is a free complex fermion and \eqref{eq:latcpfer} suggests that the complex fermions 
are coupled to each other, so we indeed have coupled free fermions in $V_{\frac{1}{\sqrt{2}}A_n}$ as desired.

We shall derive an energy-momentum tensor $L(z)$ in terms of coupled free fermions as in \eqref{eq:CFFEM}. 
First, by expanding $e^{\frac{i}{\sqrt{2}}\alpha\cdot\Phi(z)}e^{-\frac{i}{\sqrt{2}}\alpha\cdot\Phi(z)}$, we see that
\[:\psi^{\alpha}(z)\overline{\psi^{\alpha}}(z):=\frac{i}{\sqrt{2}}\alpha\cdot\partial\Phi(z).\]
Then we have
\[L^{\alpha}(z):=-\frac{1}{2}:\psi^{\alpha}(z)\partial\overline{\psi^{\alpha}}(z):=\frac{1}{2}(\frac{i}{\sqrt{2}}\alpha\cdot\partial\Phi)(z)(\frac{i}{\sqrt{2}}\alpha\cdot\partial\Phi)(z).\]
We construct the total energy-momentum tensor as
\[L(z)=\frac{2}{n+1}\sum_{\alpha\in\Delta_{+}}L^{\alpha}(z)=-\frac{1}{n+1}\sum_{\alpha\in\Delta_{+}}:\psi^{\alpha}(z)\partial\overline{\psi^{\alpha}}(z):\]
so that the conformal weight of the field $\psi^{\alpha}$ for any $\alpha\in A_n$ is indeed $\frac{1}{2}$.

Since $\frac{1}{\sqrt{2}}A_n$ is not integral, $\frac{1}{\sqrt{2}}A_n\not\subset \left(\frac{1}{\sqrt{2}}A_n\right)^*$
and we cannot expect to classify the modules by the cosets of $\frac{1}{\sqrt{2}}A_n$ in its dual as one does for $V_{\sqrt{2}A_n}$. 
However, we notice that $\frac{1}{\sqrt{2}}A_n\subset \frac{1}{2}\left(\frac{1}{\sqrt{2}}A_n\right)^*=\frac{1}{\sqrt{2}}A^*_n$, 
so we may expect the modules correspond to the coset $\left(\frac{1}{\sqrt{2}}A^*_n\right)/\left(\frac{1}{\sqrt{2}}A_n\right)$. 
If this is the case, for $\gamma\in\left(\frac{1}{\sqrt{2}}A^*_n\right)/\left(\frac{1}{\sqrt{2}}A_n\right)$, the character would be given by
\begin{equation}\label{eq:latch-scaled}
    \dfrac{1}{\eta(\tau)^{n}}\sum_{\nu\in\gamma+\frac{1}{\sqrt{2}}A_n}q^{\frac{1}{2}|\nu|^2}.
\end{equation}

We can first test our conjecture on $V_{\frac{1}{\sqrt{2}}A_1}$. Let $\alpha$ denote the root of $A_1$. 
It is obvious that there is only one complex free fermion $\psi^{\alpha}$ in $V_{\frac{1}{\sqrt{2}}A_1}$. 
We can rewrite the complex free fermion $\psi^{\alpha}$ in terms of two real free fermions by setting 
\begin{align*}
    &\chi^{\alpha}(z)=\frac{1}{\sqrt{2}}\left(\psi^{\alpha}(z)+\overline{\psi^{\alpha}}(z)\right),& \xi^{\alpha}(z)=\frac{1}{i\sqrt{2}}\left(\psi^{\alpha}(z)-\overline{\psi^{\alpha}}(z)\right),
\end{align*}
for which it can be seen that
\begin{equation*}
    \begin{tabular}{ccc}
      $\chi^{\alpha}(z)\chi^{\alpha}(w)\sim\dfrac{1}{z-w}$, &$\xi^{\alpha}(z)\xi^{\alpha}(w)\sim\dfrac{1}{z-w}$, &$\chi^{\alpha}(z)\xi^{\alpha}(w)\sim0$.
    \end{tabular}
\end{equation*}
Therefore we should expect $V_{\frac{1}{\sqrt{2}}A_1}$-modules corresponding to the NS and R sectors of two (uncoupled) free fermions. 


On the other hand, we know $\frac{1}{\sqrt{2}}A_1^*=\mathbb{Z}\left\{\frac{1}{2\sqrt{2}}\alpha\right\}$ and 
therefore we expect two modules $V_{\frac{1}{\sqrt{2}}A_1}$ and $V_{\frac{1}{2\sqrt{2}}\alpha+\frac{1}{\sqrt{2}}A_1}$, 
whose characters are, according to \eqref{eq:latch-scaled},
\begin{equation}\label{eq:A1latch}
  \begin{aligned}
  \mathrm{ch}\left[V_{\frac{1}{\sqrt{2}}A_1}\right](\tau)&=\dfrac{1}{\eta(\tau)}\sum_{n\in\mathbb{Z}}q^{\frac{1}{4}|n\alpha|^2}=
  \dfrac{1}{\eta(\tau)}\sum_{n\in\mathbb{Z}}q^{\frac{1}{2}n^2},\\
   \mathrm{ch}\left[V_{\frac{1}{2\sqrt{2}}\alpha+\frac{1}{\sqrt{2}}A_1}\right](\tau)&=\dfrac{1}{\eta(\tau)}
   \sum_{n\in\mathbb{Z}}q^{\frac{1}{4}\left|\left(n+\frac{1}{2}\right)\alpha\right|^2}=\dfrac{1}{\eta(\tau)}\sum_{n\in\mathbb{Z}}q^{\frac{1}{2}\left(n+\frac{1}{2}\right)^2}.
  \end{aligned}  
\end{equation}
Comparing \eqref{eq:A1latch} and characters of fermions $\chi_h$, with $h$ the conformal dimension of the corresponding 
representation, given in \cite{SchCFT}, we note that
\begin{align*}
     &\textstyle\mathrm{ch}\left[V_{\frac{1}{\sqrt{2}}A_1}\right]=\left( \chi_0+\chi_{\frac{1}{2}}\right)^2,&\textstyle\mathrm{ch}
     \left[V_{\frac{1}{\sqrt{2}}A_1}\right]=\left( \chi_{\frac{1}{16}}\right)^2,
\end{align*}
which can be proved by the Jacobi triple identity 
\[\prod_{k=1}^{\infty} (1-q^{k})(1-z^{-1}q^{k-\frac{1}{2}})(1-zq^{k-\frac{1}{2}})=\sum_{k\in\mathbb{Z}}(-1)^kz^kq^{\frac{1}{2}k^2}.\]

We shall proceed to investigate $V_{\frac{1}{\sqrt{2}}A_2}$. Let $\{\alpha_1,\alpha_2\}$ denote the set of simple roots of $A_2$ and set $\alpha_3=\alpha_1+\alpha_2$. 
Then we have three complex fermions $\psi^{\alpha_1}$, $\psi^{\alpha_2}$, $\psi^{\alpha_3}$ in $V_{\frac{1}{\sqrt{2}}A_2}$. We can also rewrite 
them in terms of real fermions by setting, for $i=1,2,3$,
\begin{align}\label{eq:A2realfer}
    &\chi^{i}(z)=\frac{1}{\sqrt{2}}\left(\psi^{\alpha_i}(z)+\overline{\psi^{\alpha_i}}(z)\right),&\xi^{i}(z)=\frac{1}{i\sqrt{2}}\left(\psi^{\alpha_i}(z)-\overline{\psi^{\alpha_i}}(z)\right).
\end{align}

We notice that if the values of $c_{i,j}$ are 8th root of unity with the $\pm$ signs chosen carefully, then the OPEs of $\{\chi_i,\xi_i:i=1,2,3\}$ 
are exactly the same as the OPEs for generating fields in $\widehat{\mathfrak{sl}(4)}_2/\widehat{\mathfrak{u}(1)}^3$ as given by \eqref{eq:freefer} 
and \eqref{eq:couplefer} with parameters given in \eqref{eq:sl4para}. Explicitly, we have following correspondence:
\begin{equation}\label{eq:A2sl4corrsp}
    \begin{array}{cccccc}
    \chi^1\leftrightarrow \psi^{(1)}, & \chi^2\leftrightarrow \psi^{(2)}, & \chi^3\leftrightarrow \psi^{(4)}, & \xi^1\leftrightarrow \psi^{(3)},& 
    \xi^2\leftrightarrow \psi^{(6)}, & \xi^3\leftrightarrow \psi^{(5)}.
    \end{array}
\end{equation}

Hence we expect that the $V_{\frac{1}{\sqrt{2}}A_2}$-modules are related to $\widehat{\mathfrak{sl}(4)}_2/\widehat{\mathfrak{u}(1)}^3$-modules 
in one way or another. We may investigate their relations by looking at the characters.

We first note that $\frac{1}{\sqrt{2}}A_2^*$ has a basis $\left\{\frac{1}{3\sqrt{2}}(\alpha_1-\alpha_2),\frac{1}{\sqrt{2}}\alpha_1\right\}$, 
so $\left(\frac{1}{\sqrt{2}}A_2^*\right)/\left(\frac{1}{\sqrt{2}}A_2\right)$ has three cosets that can be represented by 
\begin{equation*}
    \begin{array}{ccc}
       \gamma_0:=0,  & \gamma_1:=\frac{1}{3\sqrt{2}}(\alpha_1-\alpha_2),&\gamma_2:=\frac{2}{3\sqrt{2}}(\alpha_1-\alpha_2).
    \end{array}
\end{equation*}
The characters of $V_{\gamma_i+\frac{1}{\sqrt{2}}A_2}$ are then given by 
\begin{align}
    \mathrm{ch}\left[V_{\frac{1}{\sqrt{2}}A_2}\right](\tau)&=\eta(\tau)^{-2}\sum_{m,n\in\mathbb{Z}}q^{\frac{1}{2}(m^2+n^2-mn)},\label{eq:A2latuntwi}\\
    \mathrm{ch}\left[V_{\gamma_1+\frac{1}{\sqrt{2}}A_2}\right](\tau)=\mathrm{ch}\left[V_{\gamma_2+\frac{1}{\sqrt{2}}A_2}\right](\tau)&
    =\eta(\tau)^{-2}q^{\frac{2}{3}}\sum_{m,n\in\mathbb{Z}}q^{\frac{1}{2}(m^2+n^2-mn+2m)}.\label{eq:A2lattwi}
\end{align}
Comparing \eqref{eq:A2latuntwi} and \eqref{eq:A2lattwi} with \eqref{eq:sl4strfun}, we find that
\begin{align}
    \mathrm{ch}\left[V_{\frac{1}{\sqrt{2}}A_2}\right]&=b^{2000}_{2000}+b^{2000}_{0020}+6b^{2000}_{0101},\label{eq:A2UCPF-1}\\
     \mathrm{ch}\left[V_{\gamma_1+\frac{1}{\sqrt{2}}A_2}\right]&=2\left(b^{0101}_{2000}+3b^{0101}_{0101}\right),\label{eq:A2UCPF-2}
\end{align}
which support our conjecture on the connection between $V_{\frac{1}{\sqrt{2}}A_2}$ and $\widehat{\mathfrak{sl}(4)}_2/\widehat{\mathfrak{u}(1)}^3$.

Ideally, we expect that the $\frac{1}{8}$-twisted sector of $\widehat{\mathfrak{sl}(4)}_2/\widehat{\mathfrak{u}(1)}^3$ could 
also be recovered from the modules of $V_{\frac{1}{\sqrt{2}}A_2}$. It is seen that the shifted lattices can not lead us any further. 
Motivated by the decomposition of the $\mathbb{Z}_2$-twisted orbifold character $V^{\mathbb{Z}_2}_{\sqrt{2}A_2}$ and the fact that the 
conformal dimension of the highest weight state in the $\mathbb{Z}_2$-orbifold from the lattice $\frac{1}{\sqrt{2}}A_2$ 
would be $\frac{d}{16}=\frac{1}{8}$, we shall consider the $\mathbb{Z}_2$-orbifold of $V_{\frac{1}{\sqrt{2}}A_2}$, denoted by $V_{\frac{1}{\sqrt{2}}A_2}^{\mathbb{Z}_2}$.

 Suppose that we have a reflection map $\theta$ acts by taking the bosons $\Phi=(\phi^1,\phi^2)^T$ to $-\Phi$. We follow the orbifold 
 construction given in \cite{DGM1990orb,Dolan1996} with relaxed locality conditions whenever necessary. The orbifold characters are given by
\begin{align*}
    \text{ch}_{++}(q)=&\Tr_{\mathcal{H}_0}q^{L_0-\frac{d}{24}} ,&\text{ch}_{+-}(q)= \Tr_{\mathcal{H}_0}\theta q^{L_0-\frac{d}{24}},\\
    \text{ch}_{-+}(q)=&\Tr_{\mathcal{H}^T_0} q^{L_0+\frac{d}{48}},&\text{ch}_{--}(q)=\Tr_{\mathcal{H}^T_0} \theta q^{L_0+\frac{d}{48}},
\end{align*} 
where $\mathcal{H}$ is the fock space of the $A_2$-lattice construction, $\mathcal{H}^T(\Gamma)$ is the space built up from a 
representation $\Gamma_0$ of the gamma matrix algebra associated with $\Gamma$, and $\mathcal{H}_0$ (resp. $\mathcal{H}^T_0$) 
is the subspace of $\mathcal{H}$ (resp. $\mathcal{H}^T$) on which $\theta$ is the identity map.

The character formula $\text{ch}_{++}(q)$ gives \eqref{eq:A2latuntwi} and \eqref{eq:A2lattwi} for the untwisted sector and $\frac{1}{6}$-twisted 
sector of $V_{\frac{1}{\sqrt{2}}A_2}^{\mathbb{Z}_2}$ respectively. The reflection map $\theta$ in $\text{ch}_{+-}(q)$ gives an extra $-1$ for each simple 
root, so it gives a factor of $(-1)^{m+n}$ to the summations in \eqref{eq:A2latuntwi} and \eqref{eq:A2lattwi}, and we have
\begin{align}
   \mathrm{ch}\left[V_{\frac{1}{\sqrt{2}}A_2}^{\mathbb{Z}_2}\right]_{+-}(\tau)&=\eta(\tau)^{-2}\sum_{m,n\in\mathbb{Z}}(-1)^{m+n}q^{\frac{1}{2}(m^2+n^2-mn)},\label{eq:A2orbch-1}\\
    \mathrm{ch}\left[V_{\gamma_i+\frac{1}{\sqrt{2}}A_2}^{\mathbb{Z}_2}\right]_{+-}(\tau)&= \eta(\tau)^{-2}q^{\frac{2}{3}}\sum_{m,n\in\mathbb{Z}}(-1)^{m+n}q^{\frac{1}{2}(m^2+n^2-mn+2m)},\quad \mathrm{for}\; i=1,2.\label{eq:A2orbch-2}   
\end{align}

Geometrically, we think of the orbifold as given by $\left(A_2/\sqrt{2}\right)/\mathbb{Z}_2$. According to \cite{ItoKuniZ2string,Bagger1986}, 
the $\mathbb{Z}_2$-twisted states are located at the fixed points of the orbifold. Hence, in this case, the $\mathbb{Z}_2$-twisted 
characters are given by the squares of the characters of a free boson in the anti-periodic sector \cite{SchCFT,francesco1997conformal}, that is,
\begin{align}
      \mathrm{ch}\left[V_{\frac{1}{\sqrt{2}}A_2}^{\mathbb{Z}_2}\right]_{-+}:=[\mathrm{ch}_{-+}(\tau)]^2&=\eta(\tau)^2\eta(\tau/2)^{-2},\label{eq:A2orbch-3} \\
     \mathrm{ch}\left[V_{\frac{1}{\sqrt{2}}A_2}^{\mathbb{Z}_2}\right]_{--}:=[\mathrm{ch}_{--}(\tau)]^2&=\eta(2\tau)^2\eta(\tau/2)^2\eta(\tau)^{-4}.\label{eq:A2orbch-4} 
\end{align}
Comparing the $V_{\frac{1}{\sqrt{2}}A_2}^{\mathbb{Z}_2}$-characters \eqref{eq:A2orbch-1} - \eqref{eq:A2orbch-4} with $\widehat{\mathfrak{sl}(4)}_2$-string 
functions in \eqref{eq:sl4strfun}, we note that
\begin{align*}
    & \mathrm{ch}\left[V_{\frac{1}{\sqrt{2}}A_2}^{\mathbb{Z}_2}\right]_{+-}=b^{2000}_{2000}+b^{2000}_{0020}-2b^{2000}_{0101}, &
    \mathrm{ch}\left[V_{\gamma_i+\frac{1}{\sqrt{2}}A_2}^{\mathbb{Z}_2}\right]_{+-}=b^{0101}_{2000}-b^{0101}_{0101},\\
   &\mathrm{ch}\left[V_{\frac{1}{\sqrt{2}}A_2}^{\mathbb{Z}_2}\right]_{-+}=b^{1100}_{1100}+b^{1100}_{0011},&
   \mathrm{ch}\left[V_{\frac{1}{\sqrt{2}}A_2}^{\mathbb{Z}_2}\right]_{--}=b^{1100}_{1100}-b^{1100}_{0011},
\end{align*}
which suggest a strong connection between $V_{\frac{1}{\sqrt{2}}A_2}^{\mathbb{Z}_2}$ and $\widehat{\mathfrak{sl}(4)}_2/\widehat{\mathfrak{u}(1)}^3$.

Furthermore, we note that the UCPF of the $\frac{1}{8}$-twisted sectors of $\widehat{\mathfrak{sl}(4)}_2/\widehat{\mathfrak{u}(1)}^3$ is $4$
times $\mathrm{ch}\left[V_{\frac{1}{\sqrt{2}}A_2}^{\mathbb{Z}_2}\right]_{-+}$ \cite{BCH}.
The factor 4 here interestingly agrees with the number of fixed points of the $\mathbb{Z}_2$ action on $A_2/\sqrt{2}$.

\section{Summary}

We introduced the notion of coupled free fermions and focused on the structure of the class of coupled free fermion CFTs from 
the coset construction $\hat{\mathfrak{g}}_k/\widehat{\mathfrak{u}(1)}^n$. In particular, we studied in detail the 
examples: $\widehat{\mathfrak{sl}(3)}_2/\widehat{\mathfrak{u}(1)}^2$ and $\widehat{\mathfrak{sl}(4)}_2/\widehat{\mathfrak{u}(1)}^3$. 
We analysed their representation spaces and chiral vertex operators with extensive use of generalised commutation relations. 
We found specific bases for them so that more information about the exclusion statistics of coupled free fermions can be revealed.

We also noticed another class of coupled free fermion CFTs which are from the lattice construction based on $\frac{1}{\sqrt{2}}X_n$, 
with $X_n$ the root lattice of any simply-laced Lie algebra. We paid special attention to the example of the $\frac{1}{\sqrt{2}}A_2$ 
lattice model because of its unexpected relation to the $\widehat{\mathfrak{sl}(4)}_2/\widehat{\mathfrak{u}(1)}^3$ coset model. 
We later found out that more precisely speaking, it is the $\mathbb{Z}_2$-orbifold of $\frac{1}{\sqrt{2}}A_2$-lattice model that is intimately 
connected to $\widehat{\mathfrak{sl}(4)}_2/\widehat{\mathfrak{u}(1)}^3$ by studying the modules of $\frac{1}{\sqrt{2}}A_2$-lattice model and its orbifold. 
At this stage, we can conclude that the orbifold of the $\frac{1}{\sqrt{2}}A_2$ lattice model and the $\widehat{\mathfrak{sl}(4)}_2/\widehat{\mathfrak{u}(1)}^3$ 
coset model can be projected onto sub-sectors of each other. We obtained full correspondence between the characters of the $\mathbb{Z}_2$-orbifold of the $\frac{1}{\sqrt{2}}A_2$ 
lattice model and the characters of $\widehat{\mathfrak{sl}(4)}_2/\widehat{\mathfrak{u}(1)}^3$.
\acknowledgments
This work was partially supported by ``Tim and Margaret Bourke PhD Scholarship''. 

\appendix
\section{Proof of the spanning property of \texorpdfstring{$\mathcal{B}_1$}{B1}}\label{app:sl3}
For $\widehat{\mathfrak{sl}(3)}_2/\widehat{\mathfrak{u}(1)}^2$, due to the simple structure of $Q/2Q\cong\mathbb{Z}_2\times\mathbb{Z}_2$, 
the OPEs \eqref{eq:freefer} and \eqref{eq:couplefer}, and the generalised commutation relations \eqref{eq:gcommrel} can be summarised as
\begin{equation}\label{eq:sl3OPE}
    \begin{aligned}
    &\psi^{(i)}(z)\psi^{(i)}(w)\sim\dfrac{1}{(z-w)},\\
     &\psi^{(i)}(z)\psi^{(j)}(w)\sim\dfrac{c_{ij}\psi^{(k)}(w)}{(z-w)^{1/2}},\;\;\text{if}\;i\ne j\ne k,
    \end{aligned}
\end{equation}
and
\begin{align}
    \psi^{i}_n\psi^{i}_m+\psi^{i}_m\psi^{i}_n&=\delta_{m+n,0},&\label{eq:sl3commrel-1}\\
    \sum_{l\ge0}\binom{l-\frac{1}{2}}{l}\left(\psi^{(i)}_{m-\frac{1}{2}-l}\psi^{(j)}_{n+\frac{1}{2}+l}+
    \mu_{ij}\psi^{(j)}_{n-l}\psi^{(i)}_{m+l}\right)&=c_{ij}\psi^{(k)}_{m+n},&i\ne j\ne k, \label{eq:sl3commrel-2}
\end{align}
respectively, with constants $c_{ij}:=c_{\alpha_i,\alpha_j}$ and $\mu_{ij}:=\mu_{\alpha_i,\alpha_j}$ given in \cite{BorisPF} as follows:
\begin{equation}\label{eq:sl3parameter}
    \begin{aligned}
    &c_{ij}=\mu_{ij}c_{ji}, \; \mu_{ij}\mu_{ji}=1\\
    &\mu_{12}=\mu_{23}=\mu_{31}=x^2,\\
    &c_{12}=c_{23}=c_{31}=\frac{x}{\sqrt{2}},
    \end{aligned}
\end{equation}
where $x$ is an 8th root of unity, which we may choose to be $e^{-\frac{i\pi}{4}}$ here.

We say a state is of length $n$ if its total number of modes is $n$. We say a state in the Fock space of the untwisted 
sector of $\widehat{\mathfrak{sl}(3)}_2/\widehat{\mathfrak{u}(1)}^2$ of length $n$ is well-ordered if it is $0$ or a linear 
combination of the states in the form of \eqref{eq:sl3basis-123} of length less or equal than $n$. Let $P(n)$ be 
the statement that any state in the Fock space of the untwisted sector of $\widehat{\mathfrak{sl}(3)}_2/\widehat{\mathfrak{u}(1)}^2$ with $n$ modes can be well-ordered.

$P(1)$ is obvious since we have $\psi^{(i)}_{m}\ket{0}=0$ for any $i=1,2,3$ and $m\ge 0$. Now suppose $P(n)$ for all $n<N$, 
we intend to prove $P(N)$. With this inductive hypothesis, we can ignore the terms on the right-hand side of \eqref{eq:sl3commrel-1} 
and \eqref{eq:sl3commrel-2} in the following discussion. The exact coefficients in the generalised commutation relations are not crucial 
to this discussion either. We use the notation $\sim$ to suppress such non-crucial information and view \eqref{eq:sl3commrel-1} and \eqref{eq:sl3commrel-2} as 
\begin{align}
    &\psi^{(i)}_n \psi^{(i)}_m \sim  \psi^{(i)}_m \psi^{(i)}_n \label{eq:sl3com-sim-1},\\
    &\psi^{(i)}_{r}\psi^{(j)}_{s}\sim\sum_{l\ge 1}\psi^{(i)}_{r-l}\psi^{(j)}_{s+l}+\sum_{l\ge 0}\psi^{(j)}_{s-\frac{1}{2}-l}\psi^{(i)}_{r+\frac{1}{2}+l}\label{eq:sl3com-sim-2}.
\end{align}
We observe two facts about \eqref{eq:sl3com-sim-1} and \eqref{eq:sl3com-sim-2}:
\begin{enumerate}[label=\textbf{F.\arabic*}]
    \item They will not change the number of $\psi^{(i)}$-modes for each $i=1,2,3$.\label{fact:sl3-basis-1}
    \item They will not change the total mode number of a state. \label{fact:sl3-basis-2}   
\end{enumerate}

Let $\ket{\psi}:=\psi^{(i_N)}_{m_N}\psi^{(i_{N-1})}_{m_{N-1}}\cdots \psi^{(i_1)}_{m_1}\ket{0}$ be a state with $N$ modes. Because $P(N-1)$, 
we can assume that $\ket{\psi'}:=\psi^{(i_{N-1})}_{m_{N-1}}\cdots \psi^{(i_1)}_{m_1}\ket{0}$ is in the form of \eqref{eq:sl3basis-123} 
with $N_1$ $\psi^{(1)}$-modes, $N_2$ $\psi^{(2)}$-modes and $N_3$ $\psi^{(3)}$-modes. We have the following cases:
\begin{enumerate}[label=\textbf{C.\arabic*}]
    \item If $i_N=i_{N-1}=i$ and 
    \begin{enumerate}
        \item $m_N<m_{N-1}$, then we are done.
        \item $m_N= m_{N-s}$ for some $s\le N_i$, then by using \eqref{eq:sl3com-sim-1} repeatedly we have $\ket{\psi}\sim 0$.
        \item $m_{N-s}<m_N< m_{N-s-1}$ for some $0<s< N_i$, or $m_{N-s}<m_N\le -\sum_{j=1}^i \frac{N_j}{2}-\frac{1}{2}$ for $s=N_i$, 
        then by using \eqref{eq:sl3com-sim-1} repeatedly we have
    \[\ket{\psi}\sim \psi^{(i_{N-1})}_{m_{N-1}} \cdots\psi^{(i_{N-s})}_{m_{N-s}}\psi^{(i_N)}_{m_N}\psi^{(i_{N-s-1})}_{m_{N-s-1}}\cdots \psi^{(i_1)}_{m_1}\ket{0}\]
    which is well-ordered.
    \item $m_N> -\sum_{j=1}^i \frac{N_j}{2}-\frac{1}{2}$, then by using \eqref{eq:sl3com-sim-1} repeatedly we have
    \[\ket{\psi}\sim \psi^{(i)}_{m_{N-1}} \cdots\psi^{(i)}_{m_{N-N_i}}\psi^{(i)}_{m_N}\psi^{(i_{N-N_i-1})}_{m_{N-N_i-1}}\cdots \psi^{(i_1)}_{m_1}\ket{0}.\]
    Because of $P(N-N_i)$ and \ref{fact:sl3-basis-1}, we see that $$\psi^{(i)}_{m_N}\psi^{(i_{N-N_i-1})}_{m_{N-N_i-1}}\cdots \psi^{(i_1)}_{m_1}\ket{0}\sim 
    \sum_k\psi^{(i)}_{m^{(k)}_N}\psi^{(i_{N-N_i-1})}_{m^{(k)}_{N-N_i-1}}\cdots \psi^{(i_1)}_{m^{(k)}_1}\ket{0}=:\sum_k\ket{\psi_{(k)}}$$ with $m^{(k)}_N\le -
    \sum_{j=1}^i \frac{N_j}{2}-\frac{1}{2}$,$\forall k$. Recall that $\ket{\psi'}$ is in the form of \eqref{eq:sl3basis-123}, so we know 
    $m_{N-1}<\cdots<m_{N-N_i}\le -\sum_{j=1}^i \frac{N_j}{2}-\frac{1}{2}$. Therefore we have must have for each $k$ that $m^{(k)}_N\le 
    m_{N-1}$, or $m^{(k)}_N\ge m_{N-N_i}$, or there exists $1< s< N_i$ such that $m_{N-s}\le m^{(k)}_N\le m_{N-s-1}$. 
    Hence by using \eqref{eq:sl3com-sim-1} repeatedly we know $\psi^{(i)}_{m_{N-1}} \cdots\psi^{(i)}_{m_{N-N_i}}\ket{\psi_{(k)}}$ can be well-ordered. \label{case:sl3-1-4} 
    \end{enumerate} \label{case:sl3-1}
    \item If $i=i_N>i_{N-1}=j$ and 
    \begin{enumerate}
        \item $m_N\le-\frac{N}{2}$, then we are done.
        \item $m_N> -\frac{N}{2}$, then by \eqref{eq:sl3com-sim-2} we have     
            \begin{align*}
                 &\psi^{(i)}_{m_N}\psi^{(j)}_{m_{N-1}}\psi^{(i_{N-2})}_{m_{N-2}}\cdots \psi^{(i_1)}_{m_1}\ket{0}\tag{$\medsquare$}\\
           \sim&\sum_{l_1\ge 1}\psi^{(i)}_{m_N-l_1}\psi^{(j)}_{m_{N-1}+l_1}\psi^{(i_{N-2})}_{m_{N-2}}\cdots \psi^{(i_1)}_{m_1}\ket{0}\tag{$\medstar$}\\
           +&\sum_{l_1\ge 0}\psi^{(j)}_{m_{N-1}-\frac{1}{2}-l_1}\psi^{(i)}_{m_N+\frac{1}{2}+l_1}\psi^{(i_{N-2})}_{m_{N-2}}\cdots \psi^{(i_1)}_{m_1}\ket{0}.\tag{$\meddiamond$}
            \end{align*}
We note that because of $P(N-1)$ and \ref{fact:sl3-basis-1}, we know that for the terms in $(\meddiamond)$ we have
\begin{align*}
    &\psi^{(j)}_{m_{N-1}-\frac{1}{2}-l_1}\psi^{(i)}_{m_N+\frac{1}{2}+l_1}\psi^{(i_{N-2})}_{m_{N-2}}\cdots \psi^{(i_1)}_{m_1}\ket{0}\\
    \sim\,&\psi^{(j)}_{m_{N-1}-\frac{1}{2}-l_1}\sum_{k}\psi^{(i)}_{m^{(k)}_{N}}\psi^{(i_{N-2})}_{m^{(k)}_{N-2}}\cdots \psi^{(i_1)}_{m^{(k)}_1}\ket{0}, 
    &\mathrm{with}\;\;m^{(k)}_{N}\le -\frac{N}{2}+\frac{1}{2}.
\end{align*}
Now using \eqref{eq:sl3com-sim-2} again, we have
\begin{align*}
    &\psi^{(j)}_{m_{N-1}-\frac{1}{2}-l_1}\psi^{(i)}_{m^{(k)}_{N}}\psi^{(i_{N-2})}_{m^{(k)}_{N-2}}\cdots \psi^{(i_1)}_{m^{(k)}_1}\ket{0}\tag{$\filledsquare$}\\
    \sim& \sum_{r_1\ge 1}\psi^{(j)}_{m_{N-1}-\frac{1}{2}-l_1-r_1}\psi^{(i)}_{m^{(k)}_{N}+r_1}\psi^{(i_{N-2})}_{m^{(k)}_{N-2}}\cdots \psi^{(i_1)}_{m^{(k)}_1}\ket{0}\tag{$\filledstar$}\\
    +& \sum_{r_1\ge 0} \psi^{(i)}_{m^{(k)}_{N}-\frac{1}{2}-r_1}\psi^{(j)}_{m_{N-1}-l_1+r_1}\psi^{(i_{N-2})}_{m^{(k)}_{N-2}}\cdots \psi^{(i_1)}_{m^{(k)}_1}\ket{0}\tag{$\filleddiamond$}
\end{align*}
Since $m^{(k)}_{N}-\frac{1}{2}-r_1\le -\frac{N}{2}$ for any $r_1$ and $P(N-1)$, we see that ($\filleddiamond$) can be well-ordered. Then by using ($\filledstar$) 
repeatedly, we have  
\begin{align*}      (\filledsquare)\sim&\sum_{l_1\ge1}\sum_{l_2\ge1}\cdots\sum_{l_s\ge1}\psi^{(j)}_{m_{N-1}-\frac{1}{2}-l_1-(r_1+\cdots+r_s)}
\psi^{(i)}_{m^{(k)}_{N}+r_1+\cdots+r_s}\psi^{(i_{N-2})}_{m^{(k)}_{N-2}}\cdots \psi^{(i_1)}_{m^{(k)}_1}\ket{0}\\
        +&\text{well-ordered terms},
 \end{align*}
 where $s$ is a large enough number so that $m^{(k)}_{N}+s+\sum_{j=1}^{N-2}m^{(k)}_j\ge 0$. Hence by $P(N-1)$ and \ref{fact:sl3-basis-2}, 
 we know that $\psi^{(i)}_{m^{(k)}_{N}+r_1+\cdots+r_s}\psi^{(i_{N-2})}_{m^{(k)}_{N-2}}\cdots \psi^{(i_1)}_{m^{(k)}_1}\ket{0}$ is $0$ since it 
 has non-negative total mode number, which gives that ($\filledsquare$) and therefore $(\meddiamond)$ can be well-ordered. Similarly, 
 by using ($\medstar$) repeatedly, we see ($\medsquare$) can be well-ordered. \label{case:sl3-2b}  
    \end{enumerate} \label{case:sl3-2}   
    \item If $i=i_N<i_{N-1}=j$, then by \eqref{eq:sl3com-sim-2} we have     
            \begin{align*}
                 &\psi^{(i)}_{m_N}\psi^{(j)}_{m_{N-1}}\psi^{(i_{N-2})}_{m_{N-2}}\cdots \psi^{(i_1)}_{m_1}\ket{0}\tag{$\heartsuit$}\\
           \sim&\sum_{l_1\ge 1}\psi^{(i)}_{m_N-l_1}\psi^{(j)}_{m_{N-1}+l_1}\psi^{(i_{N-2})}_{m_{N-2}}\cdots \psi^{(i_1)}_{m_1}\ket{0}\tag{$\medcircle$}\\
           +&\sum_{l_1\ge 0}\psi^{(j)}_{m_{N-1}-\frac{1}{2}-l_1}\psi^{(i)}_{m_N+\frac{1}{2}+l_1}\psi^{(i_{N-2})}_{m_{N-2}}\cdots \psi^{(i_1)}_{m_1}\ket{0}.\tag{$\triangle$}
            \end{align*}
    Here ($\triangle$) can be well-ordered by \ref{case:sl3-2} and therefore ($\heartsuit$) can be well-ordered using ($\medcircle$) repeatedly, 
    as argued in \ref{case:sl3-2b}. \label{case:sl3-3}
\end{enumerate}
\qed





\bibliographystyle{JHEP}
\bibliography{biblio.bib}






\end{document}